# Charged Dark Energy Stars in a Finch-Skea Spacetime

Manuel Malaver[1], Rajan Iyer[2]

[1]Maritime University of the Caribbean, Department of Basic Sciences, Catia la Mar, Venezuela.
 Email: **mmf.umc@gmail.com**

[2]Environmental Materials Theoretical Physicist, Department of Physical Mathematics Sciences Engineering Project Technologies, Engineeringinc International Operational Teknet Earth Global, Tempe, Arizona, United States of America
 Email: engginc@msn.com

**Abstract:** In this paper we obtained some spherically stellar configurations that represent new models of dark energy stars specifying particular forms for gravitational potential and the electric field intensity which allows solve the Einstein-Maxwell field equations. We have chosen the metric potential proposed by Finch and Skea (1989) with the equation of state $p_r = \omega\rho$ where $p_r$ is the radial pressure, $\rho$ is the dark energy density and $\omega$ is the dark energy parameter. We found that the radial pressure, the anisotropy factor, energy density, metric coefficients, mass function, charge density are regular and well behaved in the stellar interior but the causality conditions and of strong energy are not satisfied. These models have great application in physics and cosmology due to the fact that several independent observations indicate that the universe is in a phase of accelerated expansion which can be explained by the presence of dark energy that has not been detected.
**Keywords:** Dark energy stars, Metric potential, Causality condition, Strong energy condition, Accelerated expansion.

## 1. Introduction

   Recent observational evidence as measurements of supernovas of type Ia and microwave background radiation suggest an accelerated expansion of the universe [1] and the explanation for this cosmological behavior requires assumption that a considerable part of the universe consists of a hypothetical dark energy with a negative pressure component [2], which is a cosmic fluid parameterized by an equation of state $\omega = p/\rho < -1/3$ where p is the spatially homogeneous pressure and ρ the dark energy density [1-4]. The range for which $\omega < -1$ has been denoted as phantom energy and possesses peculiar properties, such as negative temperatures and the energy density increases to infinity in a finite time, resulting in a big rip [2-4]. It also provides a natural scenario for the existence of exotic geometries such as wormholes [5-7].

   The notion of dark energy is that of a homogeneously distributed cosmic fluid and that when extended to inhomogeneous spherically symmetric spacetimes, the pressure appearing in the equation of state shows a negative radial pressure, and the tangential pressure must be determined by applying the field equations [2,3]. Lobo [3] explored several configurations, by imposing specific choices for the mass function and studied the dynamical stability of these models by applying the general stability formalism developed by Lobo and Crawford

[8]. Chan et al. [9] have proposed that the mass function is a natural consequence of the Einstein´s field equations and hence can be a core with a homogeneous energy density, described by the Lobo´s first solution [3]. Malaver and Esculpi [10] presented a new model of dark energy star by imposing specific choice for the mass function that corresponds to an increase in energy density inside of the star. Bibi et al. [4] obtained a new class of solutions of the Einstein-Maxwell field equations representing a model for dark energy stars with the equation of state $p_r=-\rho$. Malaver et al. [11] found a new family of solutions to the Einstein-Maxwell system considering a particular form of the gravitational potential $Z(x)$ and the electric field intensity with a linear equation of state that represents a model of dark energy star. Malaver and Kasmaei [12] generated a dark energy star model with a quadratic equation of state and a specific charge distribution. More recently, Malaver et. al [13] obtained new solutions of Einstein's field equations for dark energy stars within a Buchdahl spacetime by considering nonlinear electromagnetic field. According to Chan et al. [9] the denomination of dark energy is applied to fluids which violate only the strong energy condition (SEC) given by $\rho+p_r+2p_t \geq 0$ where $\rho$ is the energy density, $p_r$ and $p_t$ are the radial pressure and tangential pressure, respectively.

Recently, astronomical observations of compact objects have allowed new findings of neutron stars and strange stars that adjust to the exact solutions of the 4-D Einstein field equations; data on mass maximum, redshift and luminosity are some of the most relevant characteristics for verifying the physical requirements of these models [14]. A great number of exact models from the Einstein-Maxwell field equations have been generated by Gupta and Maurya [15], Kiess [16], Mafa Takisa and Maharaj [17], Malaver and Kasmaei [18], Malaver [19, 20], Ivanov [21] and Sunzu et al [22]. In the development of these models, several forms of equations of state can be considered [23]. Komathiraj and Maharaj [24], Malaver [25], Bombaci [26], Thirukkanesh and Maharaj [27], Dey et al. [28] and Usov [29] assume linear equation of state for quark stars. Feroze and Siddiqui [30] considered a quadratic equation of state for the matter distribution and specified particular forms for the gravitational potential and electric field intensity. Mafa Takisa and Maharaj [17] obtained new exact solutions to the Einstein-Maxwell system of equations with a polytropic equation of state. Thirukkanesh and Ragel [31] have obtained particular models of anisotropic fluids with polytropic equation of state which are consistent with the reported experimental observations. Malaver [32] generated new exact solutions to the Einstein-Maxwell system considering Van der Waals modified equation of state with polytropic exponent. Tello-Ortiz et al. [33] found an anisotropic fluid sphere solution of the Einstein-Maxwell field equations with a modified version of the Chaplygin equation of state.

The analysis of compact objects with anisotropic matter distribution is very important, because that the anisotropy plays a significant role in the studies of relativistic spheres of fluid [34-46]. Anisotropy is defined as $\Delta = p_t - p_r$ where $p_r$ is the radial pressure and $p_t$ is the tangential pressure. The existence of solid core, presence of type 3A superfluid [47], magnetic field, phase transitions, a pion condensation and electric field [29] are most important reasonable facts that explain the presence of tangential pressures within a star. Many astrophysical objects as X-ray pulsar, Her X-1, 4U1820-30 and SAXJ1804.4-3658 have anisotropic pressures. Bowers and Liang [46] include in the equation of hydrostatic equilibrium the case of local anisotropy. Bhar et al. [48] have studied the behavior of relativistic objects with locally anisotropic matter distribution considering the Tolman VII form for the gravitational potential with a linear relation between the energy density and the radial pressure. Malaver [49-50], Feroze and Siddiqui [30,51] and Sunzu et al. [22] obtained solutions of the Einstein-Maxwell field equations for charged spherically symmetric space-time by assuming anisotropic pressure. A realistic stellar model based on an ansatz of Duorah and Ray, anew analytical stellar model in general relativity, and a model of a three-layered relativistic star has been advanced to address quantum gravity astrophysics [52-54]. Recently, Malaver and Iyer [55] have analytical Equation of State models with modified Chaplygin (exotic gas that allows supersymmetric generalization, and free Tachyons with dark energy fluid having viscous generalized hybrid hypothetical substance that satisfies an exotic equation of state in the form $P = -A/\rho^{\alpha}$, where p is the pressure, $\rho$ is the density, with $\alpha =1$ and $A$ positive constant). These may be used in the description of compact objects in absence of charge as well as for the study of internal structure of strange quark stars; a strange star model may be compatible with the compact star. Energetic stars are known to have vortex action fields to churn energy and matter [56-60]. Vortex action mechanism has been modeled by breakthrough formalism examining quantum fields point model algorithmically gaging to electromagnetic fields provided stringmetrics that are associated quantum to mesoscopic to astrophysics [56-59]. In these theoretical investigations, quantum critical signal/noise density matrix values may give insight to analyze physical features associating the matter, radial pressure, density, anisotropy, gravitational potential, and energy density Schwarzschild-Einstein-Maxwell metrics.

The aim of this paper is to generate new class of solutions which represents a potential model of dark energy stars whose equation of state is $p_r = \omega\rho$ with anisotropic matter quantifiable distribution, specifying forms for the gravitational potential and the electric field intensity. We have used the ansatz proposed by Finch and Skea [52]. The systems of field equations with their derivations computations to obtain analytic solutions have been extensively detailed here to verify which are physically acceptable. We assume that the denomination of dark energy is applicable to fluids which violate the strong energy condition [9]. Organization of this article is as follows. In Section 2, we present Einstein´s field equations. In Section 3, we make a particular choice of gravitational potential $Z(x)$ that allows solving the field equations and we have obtained new models for dark energy stars consistent alone of dark matter. In Section 4, a physical analysis of the new solutions is performed. Finally in Section 5, we conclude.

## 2. Einstein-Maxwell Field Equations

We consider a spherically symmetric, static and homogeneous spacetime. In Schwarzschild coordinates the metric is given by

$$ds^2 = -e^{2\nu(r)}dt^2 + e^{2\lambda(r)}dr^2 + r^2(d\theta^2 + \sin^2\theta d\varphi^2) \tag{1}$$

where $\nu(r)$ and $\lambda(r)$ are two arbitrary functions.

The Einstein field equations for the charged anisotropic matter are given by

$$\frac{1}{r^2}\left(1-e^{-2\lambda}\right) + \frac{2\lambda'}{r}e^{-2\lambda} = \rho + \frac{1}{2}E^2 \tag{2}$$

$$-\frac{1}{r^2}\left(1-e^{-2\lambda}\right) + \frac{2\nu'}{r}e^{-2\lambda} = p_r - \frac{1}{2}E^2 \tag{3}$$

$$e^{-2\lambda}\left(\nu'' + \nu'^2 + \frac{\nu'}{r} - \nu'\lambda' - \frac{\lambda'}{r}\right) = p_t + \frac{1}{2}E^2 \tag{4}$$

$$\sigma = \frac{1}{r^2}e^{-\lambda}(r^2 E)' \tag{5}$$

where $\rho$ is the energy density, $p_r$ is the radial pressure, $E$ is electric field intensity, $p_t$ is the tangential pressure and primes denote differentiations with respect to r. Using the transformations, $x = cr^2$, $Z(x) = e^{-2\lambda(r)}$ and $A^2 y^2(x) = e^{2\nu(r)}$ with arbitrary constants A and c>0, suggested by Durgapal and Bannerji [53], the Einstein field equations can be written as

$$\frac{1-Z}{x} - 2\dot{Z} = \frac{\rho}{c} + \frac{E^2}{2c} \tag{6}$$

$$4Z\frac{\dot{y}}{y} - \frac{1-Z}{x} = \frac{p_r}{c} - \frac{E^2}{2c} \tag{7}$$

$$4xZ\frac{\ddot{y}}{y}+(4Z+2x\dot{Z})\frac{\dot{y}}{y}+\dot{Z}=\frac{p_t}{c}+\frac{E^2}{2c} \tag{8}$$

$$p_t = p_r + \Delta \tag{9}$$

$$\frac{\Delta}{c}=4xZ\frac{\ddot{y}}{y}+\dot{Z}\left(1+2x\frac{\dot{y}}{y}\right)+\frac{1-Z}{x}-\frac{E^2}{c} \tag{10}$$

$$\sigma^2 = \frac{4cZ}{x}(x\dot{E}+E)^2 \tag{11}$$

$\sigma$ is the charge density, $\Delta = p_t - p_r$ is the anisotropic factor and dots denote differentiation with respect to x. With the transformations of [53], the mass within a radius r of the sphere takes the form

$$m(x) = \frac{1}{4c^{3/2}} \int_0^x \sqrt{x}(\rho^* + E^2) dx \tag{12}$$

where $\quad \rho^* = \left(\frac{1-Z}{x} - 2\dot{Z}\right)c$

The interior metric (1) with the charged matter distribution should match the exterior spacetime described by the Reissner-Nordstrom metric:

$$ds^2 = -\left(1-\frac{2M}{r}+\frac{Q^2}{r^2}\right)dt^2 + \left(1-\frac{2M}{r}+\frac{Q^2}{r^2}\right)dr^2 + r^2(d\theta^2 + \sin^2\theta d\varphi^2) \tag{13}$$

where the total mass and the total charge of the star are denoted by $M$ and $q^2$, respectively. The junction conditions at the stellar surface are obtained by matching the first and the second fundamental forms for the interior metric (1) and the exterior metric (14).

In this paper, we assume the following equation of state

$$p_r = \omega\rho \tag{14}$$

where $\omega$ is the dark energy parameter.

## 3. A New Class of Models

In order to solve the Einstein field equations, we have chosen specific forms for the gravitational potential $Z(x)$ and the electrical field intensity E. Following Finch and Skea [52] and Lighuda et al.[54] we have taken the forms, respectively

$$Z(x) = \frac{1}{1+ax} \qquad (15)$$

$$\frac{E^2}{2c} = \frac{ax}{1+ax^2} \qquad (16)$$

where $a$ is a real constant. The metric potential is regular at the origin and well behaved in the interior of the sphere. The electric field is finite at the center of the star and remains continuous in the interior.

Substituting (15) and (16) in (7) we obtain

$$\rho = c\left[\frac{3a+a^2x}{(1+ax)^2} - \frac{ax}{1+ax^2}\right] \qquad (17)$$

Replacing (17) in (14), we have for the radial pressure

$$p_r = \omega c\left[\frac{3a+a^2x}{(1+ax)^2} - \frac{ax}{1+ax^2}\right] \qquad (18)$$

Using (17) in (12), the expression of the mass function is

$$M(x) = \left(\frac{1}{a}\right)^{1/4}\frac{\sqrt{2}}{8\sqrt{c}}\left[\frac{1}{2}\ln\left(\frac{x+\left(\frac{1}{a}\right)^{1/4}\sqrt{2x}+\sqrt{\frac{1}{a}}}{x-\left(\frac{1}{a}\right)^{1/4}\sqrt{2x}+\sqrt{\frac{1}{a}}}\right) + \arctan\left(\frac{\sqrt{2x}}{\left(\frac{1}{a}\right)^{1/4}}+1\right) + \arctan\left(\frac{\sqrt{2x}}{\left(\frac{1}{a}\right)^{1/4}}-1\right)\right] - \frac{\sqrt{x}}{2\sqrt{c}(1+ax)} \qquad (19)$$

With (15) and (16) in (11), the charge density is

$$\sigma^2 = \frac{2ac^2\left[(1-ax^2)\sqrt{1+ax^2} + 2(1+ax^2)\sqrt{2acx}\right]^2}{(1+ax)(1+ax)^4} \tag{20}$$

With (15), (16) and (17), the eq. (8) becomes

$$\frac{4}{1+ax}\frac{\dot{y}}{y} = \frac{\omega(3a+a^2x)}{(1+ax)^2} - \frac{(\omega+1)ax}{1+ax^2} + \frac{a}{1+ax} \tag{21}$$

Integrating (21), we obtain

$$y(x) = c_1(1+ax^2)^A (1+ax)^B e^{\frac{(\omega+1)\sqrt{a}\arctan(x\sqrt{a})}{4}} \tag{22}$$

where

$$A = -\frac{\omega+1}{8} \quad \text{and} \quad B = \frac{\omega}{2}$$

The metric functions can be written as

$$e^{-2\lambda(r)} = 1+ax \tag{23}$$

$$e^{2\nu(r)} = A^2 c_1^2 (1+ax^2)^{2A}(1+ax)^{2B} e^{\frac{(\omega+1)\sqrt{a}\arctan(x\sqrt{a})}{2}} \tag{24}$$

The anisotropy factor $\Delta$ is given by for

$$\Delta = \frac{4xc}{1+ax}\left[\frac{(4A^2a^2 - 4Aa^2)x^2 + (\omega+1)Aa^2x}{(1+ax^2)^2} + \frac{2aA}{1+ax^2} + \frac{4ABa^2x}{(1+ax^2)(1+ax)}\right.$$
$$\left. + \frac{(B^2-B)a^2}{(1+ax)^2} + \frac{(\omega+1)Ba^2}{2(1+ax)(1+ax^2)} - \frac{(\omega+1)a^2}{2(1+ax^2)^2} + \frac{(\omega+1)^2 a^2}{16(1+ax^2)^2}\right]$$
$$- \frac{ac}{(1+ax)^2}\left[1 + \frac{4Aax^2}{1+ax^2} + \frac{2xaB}{1+ax} + \frac{(\omega+1)xa}{2(1+ax^2)}\right] + \frac{ac}{1+ax} - \frac{2acx}{1+ax^2}$$

$$\tag{25}$$

## 4. Conditions of Physical Acceptability

For a model to be physically acceptable, the following conditions should be satisfied [4,41]:

(i) The metric potentials $e^{2\lambda}$ and $e^{2\nu}$ assume finite values throughout the stellar interior and are singularity-free at the center $r=0$.

(ii) The energy density $\rho$ should be positive and a decreasing function inside the star.

(iii) The radial pressure also should be positive and a decreasing function of radial parameter but for negative pressure this condition is not satisfied.

(iv) The density gradient $d\rho/dr \leq 0$ for $0 \leq r \leq R$.

(v) The anisotropy is zero at the center $r=0$, i.e. $\Delta(r=0) = 0$.

(vi) Any physically acceptable model must satisfy the causality condition, that is, for the radial sound speed $v_{sr}^2 = \dfrac{dp_r}{d\rho}$, we should have $0 \leq v_{sr}^2 \leq 1$ but the dark energy case this condition nor is it satisfied.

(vii) The consideration of dark energy is applicable only to fluids that violate the strong energy condition.

(viii) The charged interior solution should be matched with the Reissner–Nordström exterior solution, for which the metric is given by the equation (13).

The conditions (ii) and (iv) imply that the energy density must reach a maximum at the centre and decreasing towards the surface of the sphere.

## 5. Physical Analysis the New Models

For the new solutions, metric potentials $e^{2\lambda}$ and $e^{2\nu}$ have finite values and remain positive throughout the stellar interior. At the center $e^{2\lambda(0)} = 1$ and $e^{2\nu} = A^2 c_1^2$. We show that in $r=0$ $\left(e^{2\lambda(r)}\right)'_{r=0} = \left(e^{2\nu(r)}\right)'_{r=0} = 0$ and this makes is possible to verify that the gravitational potentials are regular at the center.

The energy density is positive and well behaved between the center and the surface of the star. In the center $\rho(r=0) = 3ac$ and $p_r(r=0) = 3\omega ac$, therefore the energy density will be non-negative in $r=0$ and $p_r(r=0) < 0$. In the surface of the star $r=R$ we have $p_r(r=R) = 0$ and $R = \dfrac{\sqrt{2ac\left(1 - a - \sqrt{a^2 - 14a + 1}\right)}}{2ac}$

For the density gradient inside the stellar interior, we obtain

$$\frac{d\rho}{dr} = \frac{2a^2c^2r}{(1+acr^2)^2} - \frac{4ac^2r(3a+a^2cr^2)}{(1+acr^2)^3} - \frac{2ac^2r}{1+ac^2r^4} + \frac{4a^2c^4r^5}{(1+ac^2r^4)^2} \tag{26}$$

On the boundary $r=R$, the solution must match the Reissner–Nordström exterior space–time as:

$$ds^2 = -\left(1 - \frac{2M}{r} + \frac{Q^2}{r^2}\right)dt^2 + \left(1 - \frac{2M}{r} + \frac{Q^2}{r^2}\right)^{-1} dr^2 + r^2\left(d\theta^2 + \sin^2 d\varphi^2\right)$$

and therefore, the continuity of $e^\nu$ and $e^\lambda$ across the boundary $r=R$ is

$$e^{2\nu} = e^{-2\lambda} = 1 - \frac{2M}{R} + \frac{Q^2}{R^2} \tag{27}$$

Then for the matching conditions, we obtain:

$$\frac{2M}{R} = \frac{acR^2(1+2cR^2)}{1+acR^2} \tag{28}$$

In the figures 1, 2, 3, 4 and 5 are represented the dependence of, $e^{2\lambda}$, $\rho$, $\frac{d\rho}{dr}$, $\sigma^2$ and $M(r)$ with the radial coordinate. In all the cases was it has been considered $R=1.8$ km, $a=0.02$, $c=1$.

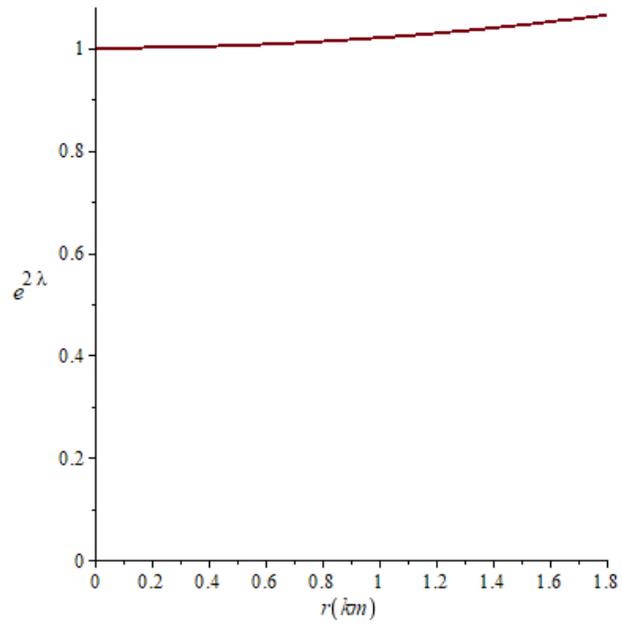

**Figure 1**. *Metric function $e^{2\lambda}$ against radial coordinate with a=0.02 and c=1.*

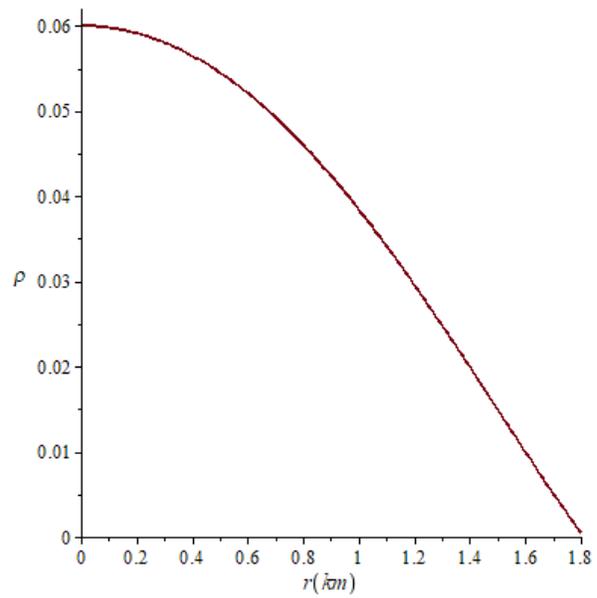

**Figure 2**. *Energy density against radial coordinate with a=0.02 and c=1.*

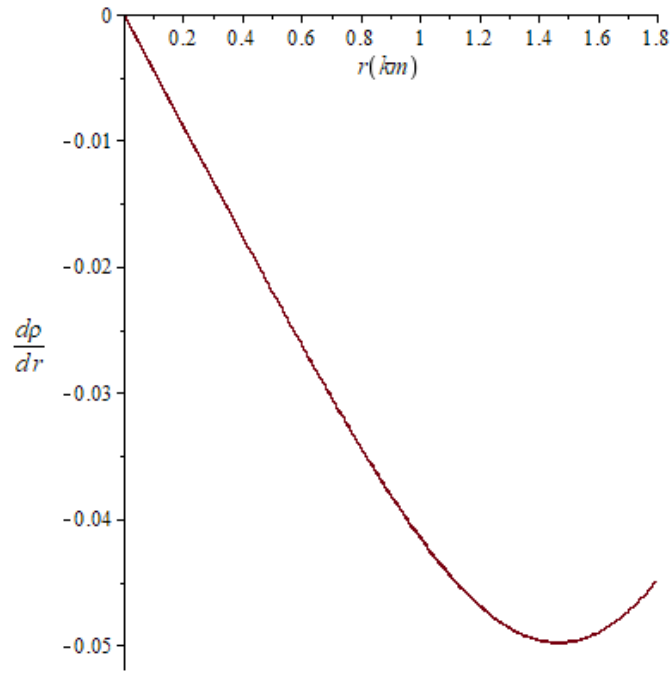

**Figure 3**. *Density gradient against radial parameter with a=0.02 and c=1.*

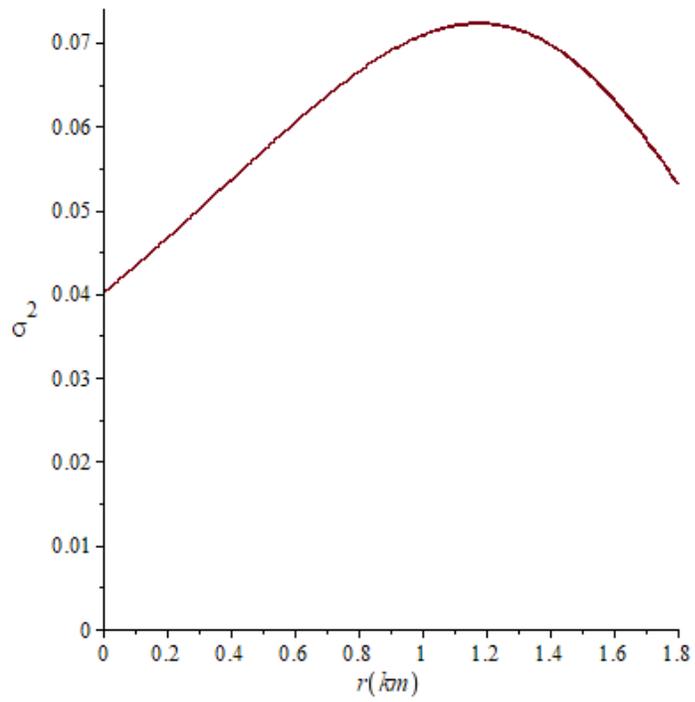

**Figure 4**. *Charge density against radial parameter with a=0.02 and c=1.*

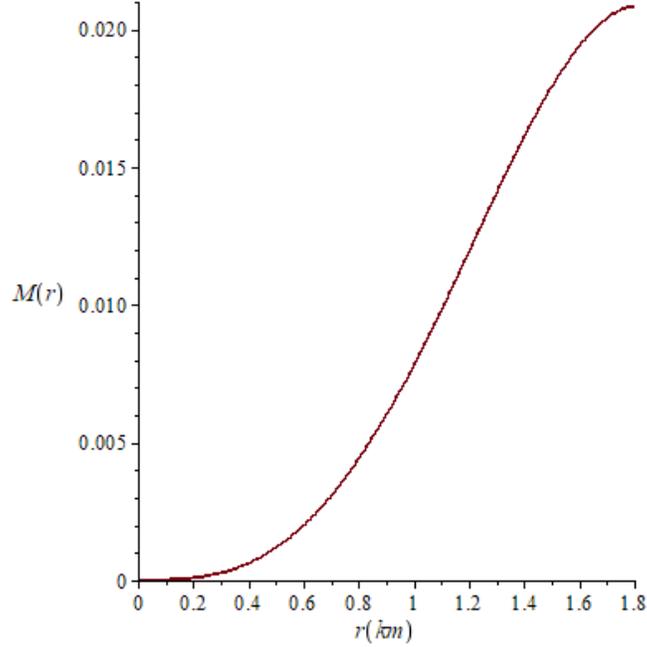

**Figure 5.** *Mass function against radial coordinate with a=0.02 and c=1*

In Figure 1 the metric potential $e^{2\lambda}$ in is a continuously growing function inside the star. The energy density remains positive, continuous and is monotonically decreasing function throughout the stellar interior as noted in the Figure 2. The radial variation of energy density gradient has been shown in Figure 3, in which it is observed that $\frac{d\rho}{dr} < 0$. In the Figure 4 the charge density is a continuously decreasing function, reaches a maximum and then decreases inside the star. In Figure 5, the mass function is continuous, increasing, takes finite values and well behaved in the stellar interior.

The Figures 6,7, 8 and 9 show the dependence of $e^{2\nu}$, $p_r$, anisotropy $\Delta$ and strong energy condition (SEC) respectively with the radial parameter for different values of $\omega$. In all the cases it has been considered *R= 1.8 Km, a=0.02* and *c = 1*.

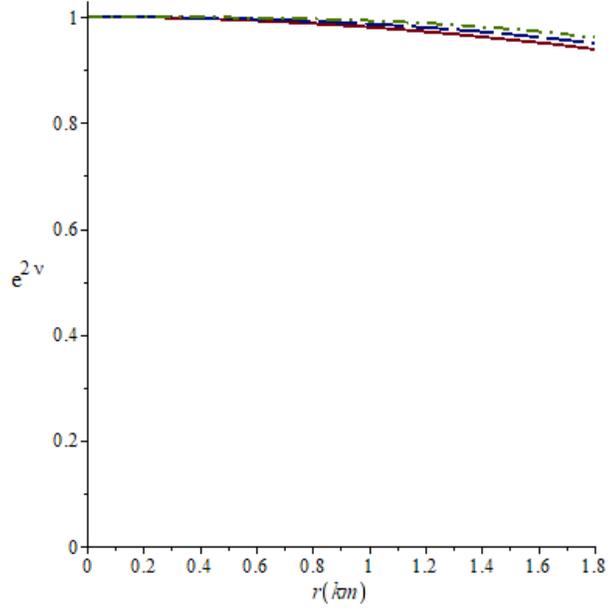

**Figure 6**. *Metric function* $e^{2v}$ *against radial coordinate for* $\omega=-1$ *(solid line);* $\omega=-0.75$ *(long-dash line):* $\omega=-0.5$ *(dashdot line). In all the cases a=0.02 and c=1.*

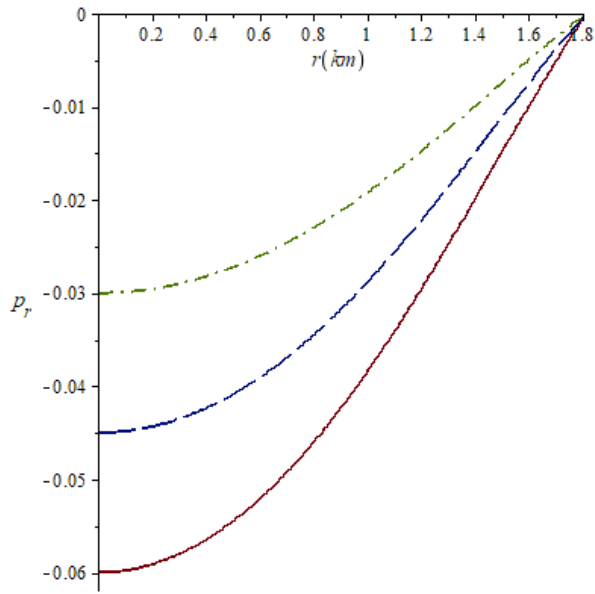

**Figure 7.** *Radial pressure against radial coordinate for* $\omega=-1$ *(solid line);* $\omega=-0.75$ *(long-dash line):* $\omega=-0.5$ *(dashdot line). In all the cases a=0.02 and c=1.*

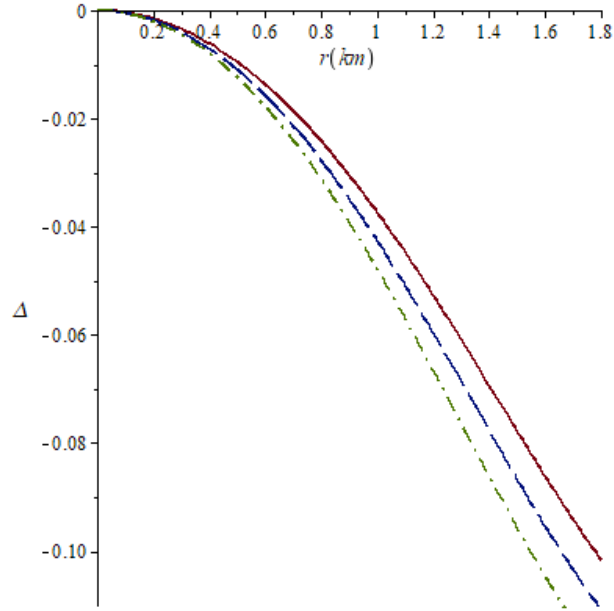

**Figure 8.** Anisotropy against radial coordinate for ω=-1 (solid line); ω=-0.75 (long-dash line): ω=-0.5 (dashdot line). In all the cases a=0.02 and c=1.

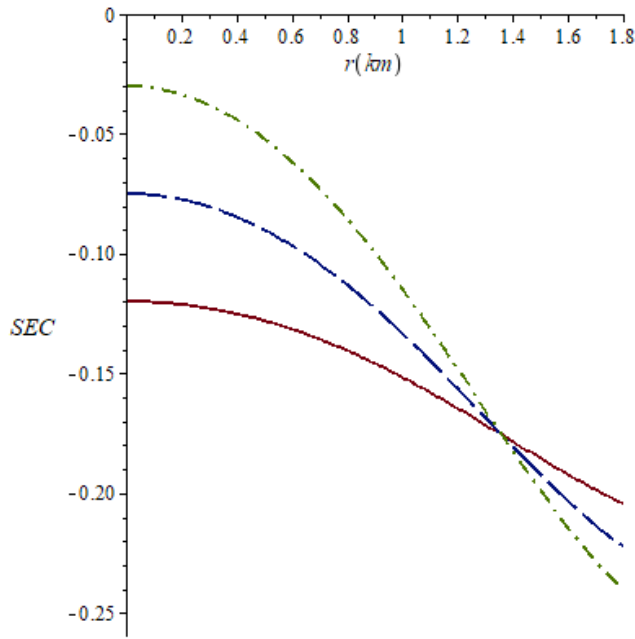

**Figure 9.** SEC against radial coordinate for ω=-1 (solid line); ω=-0.75 (long-dash line): ω=-0.5 (dashdot line). In all the cases a=0.02 and c=1.

In Figure 6 the metric potential $e^{2\nu}$ is continuous, well behaved and has a slight increase with an increase in the values of $\omega$. The radial pressure is negative and not a decreasing function of the radial parameter, but takes lower values when $\omega$ is increased as shown in

Figure 7. The anisotropic factor is plotted in Figure 8 and it shows that vanishes at the centre of the star, i.e. *Δ(r=0) =0*. We can also note that Δ admits lower values with a growth of *ω*. The Figure 9 shows that the strong energy condition is violated for all *ω* values considered.

## 6. Conclusion

In this paper we have found new class of solutions which represents a model for dark energy stars with a gravitational potential proposed for Finch and Skea [52]. The radial pressure, energy density, anisotropy, mass function, charge density and all the coefficients of the metric behaves well inside the stellar interior and are free of singularities. In this model the consideration of dark energy star is applied only to the cases where parameter *ω* not satisfy the strong energy condition. The obtained solutions match smoothly with the exterior of the Reissner–Nordström spacetime at the boundary *r=R,* because matter variables and the gravitational potentials of this work are consistent with the physical analysis of these stars. The new models satisfy all the requirements for a compact negative energy stellar object and may be used to model relativistic configurations in different astrophysical scenes.

Quantum astrophysical signal/noise density matrix with prime factorized magic square symmetry mechanistic processes have powerful methodology to characterize charged dark energy Star systems that essentially possess characteristics of Finch-Skea spacetime within vacuum multiverse. We believe that critical quantum signal/noise density matrix values may provide astrophysics physical features associating the matter, radial pressure, density, anisotropy, gravitational potential, and energy density metrics. Physics conjectures are underway proceeding to metrically gaging with unitarization to achieve dimensionless quantities with analytical physical solutions observationally. Gage discontinuity dissipative physics, then will get modified to come up with observables that are measurable to resolve questions of mass-function, fields, dark matter, as well as singularity; many inconsistencies will also get resolved by quantum astrophysical transforms.